\documentclass{sigchi}

\toappear{
Permission to make digital or hard copies of part or all of this work for personal or classroom use is granted without fee provided that copies are not made or distributed for profit or commercial advantage and that copies bear this notice and the full citation on the first page. Copyrights for third-party components of this work must be honored. For all other uses, contact the Owner/Author. \\
Copyright is held by the owner/author(s). \\
{\emph{CHI 2017}}, May 06--11, 2017, Denver, CO, USA\\
ACM 978-1-4503-4655-9/17/05.\\
\url{http://dx.doi.org/10.1145/3025453.3025847}
}

\usepackage{balance}  
\usepackage{graphics} 
\usepackage{times}    
\usepackage[hyphens]{url}      
\usepackage{microtype} 
\usepackage{array}
\usepackage[USenglish]{babel}
\usepackage{booktabs}
\usepackage{gensymb}
\usepackage{caption}
\usepackage{subcaption}
\usepackage{multirow}

\usepackage{ucs}
\usepackage[utf8x]{inputenc}

\makeatletter
\def\url@leostyle{%
  \@ifundefined{selectfont}{\def\UrlFont{\sf}}{\def\UrlFont{\small\bf\ttfamily}}}
\makeatother
\newenvironment{itemize*}%
  {\begin{itemize}%
    \setlength{\itemsep}{2pt}%
    \setlength{\parskip}{0pt}}%
  {\end{itemize}}

\def\pprw{8.5in}
\def\pprh{11in}

\graphicspath{{figures/}}

\usepackage{tikz}
\newcommand*\circled[1]{\tikz[baseline=(char.base)]{
  \node[shape=circle,draw,inner sep=1pt] (char) {#1};}}

\usepackage[pdftex]{hyperref}
\hypersetup{
pdftitle={Scratch Community Blocks: Supporting Children as Data Scientists},
pdfauthor={Sayamindu Dasgupta and Benjamin Mako Hill},
pdfkeywords={data science,learning,computers and children,creativity support tools,social computing and social navigation,block-based programming},
pdfstartview={FitH},
colorlinks,
citecolor=black,
filecolor=black,
linkcolor=black,
urlcolor=black,
breaklinks=true,
}

\usepackage[all]{hypcap}  

\begin{document}

\title{Scratch Community Blocks:\\Supporting Children as Data Scientists}

\numberofauthors{2}
\author{
  \alignauthor Sayamindu Dasgupta$^{* \dagger}$\\
    \affaddr{$^{*}$MIT Media Lab}\\
    \affaddr{Cambridge, MA 02142}\\
    \email{sayamindu@media.mit.edu}\\
  \alignauthor Benjamin Mako Hill$^{\dagger}$\\
    \affaddr{$^{\dagger}$University of Washington}\\
    \affaddr{Seattle, WA, 98195}\\
    \email{\{makohill, sdg1\}@uw.edu}
}


\maketitle

\begin{abstract}
In this paper, we present \emph{Scratch Community Blocks}, a new system that enables children to programmatically access, analyze, and visualize data about their participation in \emph{Scratch}, an online community for learning computer programming. At its core, our approach involves a shift in who analyzes data: from adult data scientists to young learners themselves. We first introduce the goals and design of the system and then demonstrate it by describing example projects that illustrate its functionality. Next, we show through a series of case studies how the system engages children in not only representing data and answering questions with data but also in self-reflection about their own learning and participation.
\end{abstract}

\keywords{
	data science; learning; computers and children; creativity support tools; social computing and social navigation; block-based programming
}

\category{K.3.2}{Computers and Education}{Computer and Information Science Education}
\category{H.5.2}{Information Interfaces and Presentation (e.g., HCI)}{User Interfaces}

\section{Introduction}

Widespread use of social media and digital learning platforms by children has led to the creation of massive sets of observational data that describe the ways that young people interact, socialize, and learn \cite{bienkowski_enhancing_2012}. In hundreds of studies using data collected from a variety of platforms and contexts, researchers have answered a wide variety of research questions about youth, influencing educational policy and instructional methods \cite{willcox_online_2016}. Although very few common threads can be drawn across the large and diverse body of ``data science'' that studies uses of sociotechnical systems by children, one common trait is that the collecting, displaying, and visualizing of data is done by adult analysts, designers, and policymakers. In most cases, this group is also charged with making sense of and acting upon analyzed data. In this process, children are the object of analysis; their role is to generate data by using the system.

\begin{figure}[t] 
\includegraphics[width=0.7\columnwidth]{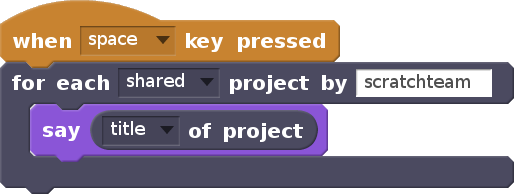}
\centering
\caption{A Scratch script, consisting of blocks from the \emph{Scratch Community Blocks} system. When the ``space'' key is pressed, this script iterates through all the shared Scratch projects by the user ``scratchteam,'' and during each iteration, a graphical object on the screen says the title of the currently selected project through a visual speech bubble.}
\label{fig:intro-script}
\end{figure}

We imagine a different approach to data science and education in which young people use data to ask and answer their own questions. Our approach is inspired by constructionism \cite{papert_mindstorms:_1980}, a theory of learning put forward by Seymour Papert and others where learners are understood to be driven by their own interests to \emph{construct} knowledge instead of passively acquiring it from a teacher. Building on Piaget's theory of constructivism \cite{piaget_genetic_1970}, constructionism theorizes the learner as an active builder of knowledge and adds the idea that knowledge building occurs ``especially felicitously in a context where the learner is consciously engaged in constructing a public entity'' \cite{papert_situating_1991}.

In this paper, we present a system called \emph{Scratch Community Blocks} that is designed to give the 15 million users of the \emph{Scratch} online community the ability to programmatically analyze data from the community itself---an ability that has previously been the exclusive domain of data scientists and engineers. The system enables community members to create and share their own visualizations and analytics tools. We begin with a brief description of a number of related systems and of Scratch. Next, we describe the motivation behind and the design of \emph{Scratch Community Blocks} and provide several short sample projects as illustrative examples. To demonstrate the range of possibilities introduced by the system, we describe a series of projects created by children using the system as part of a beta test. We discuss the technical implementation of our systems as well as several limitations of our design. We conclude by connecting back our work to some of the core aspirations of constructionist theory.

\section{Related Work}

The design of \emph{Scratch Community Blocks} is informed by a number of previous systems designed to support personal analytics, social media analytics, data science, and learning. A number of systems for data-mediated reflection and personal analytics fall into the genre of ``quantified self'' tools \cite{nafus_quantified:_2006}. These include commercially available systems (e.g.,~systems that visualize data from wearable fitness trackers) as well as systems produced by peer production \cite{benkler_wealth_2006} (e.g.,~blood glucose level analysis \cite{beck_ipancreas_2013}). Commercial systems are typically preprogrammed with limited customizability. Peer-produced tools are generally ``hackable'' but tend to be designed for experienced programmers. There is growing interest in engaging the quantified self movement in educational contexts \cite{lee_quantified_2013}.
Additionally, a number of tools exist to help end users analyze online social media systems and social networks. For example, there are widely used systems that visualize users' social connections using data from Facebook \cite{wolfram_research_wolfram|alpha_2015} and email archives \cite{viegas_visualizing_2006, jagdish_immersion_2014}. Email archives have also been used as a source of reflection with the MUSE system by Hangal et al.~\cite{hangal_muse:_2011}.

In learning technologies, the term ``dashboards'' is often used to describe systems that support the visualization and analysis of learning data. Verbert et al.~\cite{verbert_learning_2013} review 15 dashboards across a wide range of characteristics including target audience and type of data tracked. In some of these systems, data is used to inform educators about student progress with information visualization on who is progressing as expected and who is stuck (e.g.,~\cite{poldoja_edufeedr_2010}). In some cases, dashboards are presented to learners so that they can reflect. For example, the dashboard provided by the Khan Academy platform gives learners a sense of how much time they have spent on a given course module \cite{khan_academy_introducing...learning_2013}. Beyond dashboards, Rivera-Pelayo et al.~\cite{rivera-pelayo_applying_2012} presented a framework that supports reflection in informal learning through mechanisms derived from the quantified self community. Within Scratch, \emph{Jots} was an experimental system that supported Scratch users in creating brief updates or ``jots'' as they worked through their projects \cite{rosenbaum_jots_2009}.

There is also growing interest in data science tools for youth. A comparatively early tool to support learning with computation and data was \emph{Tinkerplots}, a visualization and modeling system that can be used for developing statistical reasoning skills \cite{fitzallen_developing_2010}. \emph{iSense} is a hardware toolkit and an associated web-based collaboration system that allows young learners to collect and visualize data \cite{martin_isense:_2010}. In a more specialized context, Van Wart and Parikh created \emph{Local Ground}, a system that enables youth to gain fluency with geographical information systems \cite{van_wart_increasing_2013}. \emph{BlockyTalky} \cite{shapiro_blockytalky:_2016} is a block-based visual language that enables young programmers to use sensing data from the physical world to build interactive programs.  \emph{DataSnap} is an extension to the block-based language \emph{Snap}, similar in some respects to Scratch, which can fetch and analyze data from online sources \cite{hellmann_datasnap:_2015}. \emph{DataBasic.io} is a suite of data-literacy tools, designed to used by novices in a variety of contexts \cite{bhargava_designing_2015}.

In addition to these systems, educators have frequently turned to more traditional systems to introduce data science to novices. These systems include spreadsheets such as \emph{Microsoft Excel}, visualization tools such as \emph{ManyEyes} \cite{viegas_manyeyes:_2007}, data management web-services such as \emph{Google Fusion Tables} \cite{gonzalez_google_2010}), and mainstream programming languages (e.g.,~\cite{community_data_science_collective_community_2015}, \cite{catrambone_answering_2012}).

Compared to these systems and approaches, \emph{Scratch Community Blocks} provides a unique combination of affordances; it opens up a significant amount of space for open-ended, user-driven exploration within data science, while ensuring a comparatively low barrier to entry. We know of no other system that is specifically designed for children to engage in interest-driven programmatic analysis and visualization of their own learning and social data. Like many previous systems, our work is informed by research in the learning sciences and statistics education research that has examined how children and youth establish relationships between data and context \cite{ben-zvi_childrens_2016} and how they interpret and reason about complex information visualizations \cite{laina_distributions_2016}.

\section{Scratch} 

\emph{Scratch Community Blocks} is a new system built within the \emph{Scratch} programming language \cite{resnick_scratch:_2009}. Scratch is a visual, block-based programming language designed for children and youth aged 8--16. Programs in Scratch (called ``projects'') are constructed by dragging and dropping visual blocks together. Each visual block can be thought of as a programming primitive that determines the behavior of on-screen graphical objects called ``sprites.'' For example, a script constructed out of Scratch programming blocks in Figure \ref{fig:intro-script} will make a cat sprite sequentially say the title of all the Scratch projects shared by the ``scratchteam'' user in response to a key press.

Scratch is situated within an online community where members can share projects \cite{monroy_hernandez_scratchr:_2007}. Users are encouraged to socialize around projects through commenting, bookmarking, showing appreciation, etc. Another affordance of the Scratch community is the ability to remix existing projects. Every project shared in the community is associated with a view source (``See Inside'') button, and community members are encouraged to build on and extend their peers' work. Seen through the lens of constructionism, Scratch projects are instances of ``public entities'' that are both individually and collaboratively created and shared. 

\section{Children as Data Scientists}

The central design goal of \emph{Scratch Community Blocks} is the idea of \emph{children as data scientists}. That said, it would be incorrect to describe the goal of our approach as merely shifting the ability to analyze data about youth activity online from adults to children. Just as the design of Logo by Papert and others \cite{papert_mindstorms:_1980, abelson_turtle_1986} was fundamentally shaped to support details of the constructionist theory of learning---rather than merely part of an attempt to turn children into programmers---the shift in the locus of data analysis from adults to children is only one visible feature of our approach as designers. Beyond this obvious shift lie two broad goals that parallel key features of successful constructionist learning toolkits in general. Those features are the ability to support learning through making in a social context and the ability to support self-reflection and learning about learning.

To achieve the first goal, our system aims to foster \emph{new pathways to learning data science} by enabling data science learning experiences that are discovery-driven, built around building, personally meaningful, and engaging for participants. For the second goal, the system strives to enable learners to \emph{reflect with data about their own behavior and thought processes}. The most effective constructionist systems can encourage young users to engage in self-reflection in ways that prompt them to ``embark on an exploration about how they themselves think'' \cite{papert_mindstorms:_1980}. Our goals represent our pathway, as designers, toward this central aspiration. 

\subsection{New pathways to learning data science}

Although data science is cited as an increasingly important skill or literacy, there is no general consensus on what it constitutes. We adopt one common definition that uses the term to describe a set of practices at the intersection of substantive question asking, mathematical analysis, and computing skills \cite{conway_data_2010, johnstone_data_2014}. Critically, we treat programming as a core component of data science. In contrast to most existing tools aimed at introducing youth to data analysis where visualization is central, we focus on programming as the primary way to engage with data. In our vision, programming is important because it lets the learner ask questions or conduct explorations that we as designers may not have thought of. In their discussion of the design of construction kits for children, Resnick and Silverman state that ``a little bit of programming goes a long way'' in that it opens the possibilities of a vast potential design space \cite{resnick_reflections_2005}.

Self-driven question setting also aligns well with the constructionist approach to learning that frames our approach. The core feature of any constructionist system is the ability to support learning that is personally relevant and meaningful. As a design criterion for toolkits in constructionism, this quality is referred to as ``appropriability,'' \cite{papert_mindstorms:_1980} which is described as constituting three core principles: \emph{continuity}, \emph{power}, and \emph{cultural resonance}. The continuity principle states that the topic being learned should be continuous with the prior personal knowledge of the learner. The power principle suggests that learners should be empowered to achieve new creative possibilities. The principle of cultural resonance seeks to ensure that the topic is seen as valuable in a larger social and cultural context.

\emph{Scratch Community Blocks} is designed to satisfy these principles. First, it is situated in the context of the Scratch community, and information accessed through the system is data that the users of the system and their peers have created. For the system's users (participants in the Scratch community), this creates \emph{continuity} between what users already know about the community and about programming with Scratch and the data and tools they have access through \emph{Scratch Community Blocks}. Additionally, the system builds on the visual drag-and-drop programming paradigm that Scratch users are familiar with. The system is designed to be \emph{powerful} in the sense that it provides programmatic access to data, allowing users to analyze and visualize information in ways that were unanticipated by the designers and are independent of the affordances provided by the Scratch website. Finally, given the significant amount of excitement around the emerging discipline of data science \cite{podesta_big_2014}, it is not unfair to say that the topic is seen as valuable and relevant in present society. Within the Scratch community itself, there is significant enthusiasm and excitement about the possibility of doing data analysis using Scratch data. In both senses, the system can be described as \emph{culturally resonant}.

\subsection{Reflection on learning and social participation}

Apart from constructionism, a number of theories of education highlight reflection as an important component of learning. Bruner \cite{bruner_actual_1986}, for example, theorizes that with reflection, learners ``distance'' themselves to reach a ``higher ground'' of abstraction, thereby gaining new perspectives on what has been learned. Boud, Keogh and Walker \cite{boud_reflection:_1985} propose a model of reflection in which emotion is a component and where behavior change is an important outcome of reflection. In this model, reflection requires individuals to ``recapture their experience, think about it, mull it over and evaluate it.''

Many models of reflection focus on relatively short temporal scales. For example, in Resnick's creative learning spiral model \cite{resnick_sowing_2008}, reflection is seen as the final step in one iteration of a creative activity. Schön \cite{schon_reflective_1983} differentiates between ``reflection-on-action'' and ``reflection-in-action.'' In reflection-on-action, the process of reflection happens after the action has been performed, while in reflection-in-action, the reflective process occurs hand-in-hand with the activity. The \emph{Scratch Community Blocks} system seeks to allow children to revisit and analyze data about themselves in ways that encourages reflection after the fact, but at both longer and short time scales.

\section{Design}

The \emph{Scratch Community Blocks} system is tightly integrated with the Scratch interface and community, extends the list of programming primitives in Scratch, and uses the same visual block-based drag-and-drop editing paradigm as the core Scratch language. In addition to using a block-base editing paradigm, we tried to make \emph{Scratch Community Blocks} consistent with the rest of the Scratch programming environment by (i) using programming constructs already known to Scratch users (e.g.,~``accessor'' blocks, which we describe below) and by (ii) avoiding a large number of blocks or block parameters, as Scratch has traditionally avoided having complex blocks, or a large number of blocks, to avoid intimidating novice users \cite{dasgupta_extending_2015}. This approach prevented us from considering alternate paradigms of interacting with data programmatically, such the read-eval-print loop (REPL) that is often seen in data-focused programming languages. One design that we considered involved the use of the list data structure in Scratch. We ended up not using this approach because research has shown that list are used infrequently in Scratch \cite{aivaloglou_how_2016} and also because lists are not ``first class'' (i.e.,~they cannot be used as input to an outer block or inserted into another list).

\begin{figure}[h!]
\includegraphics[width=0.45\columnwidth]{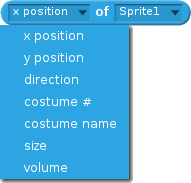}
\centering
\caption{Accessor method block for Scratch sprites, with a dropdown to select the appropriate property. The shape of the block indicates it is a ``reporter'' block, and that it can be used as an input to another blocks (e.g.,~the ``say'' block).}
\label{fig:accessor}
\end{figure}

\begin{figure}[h!]
\includegraphics[width=0.75\columnwidth]{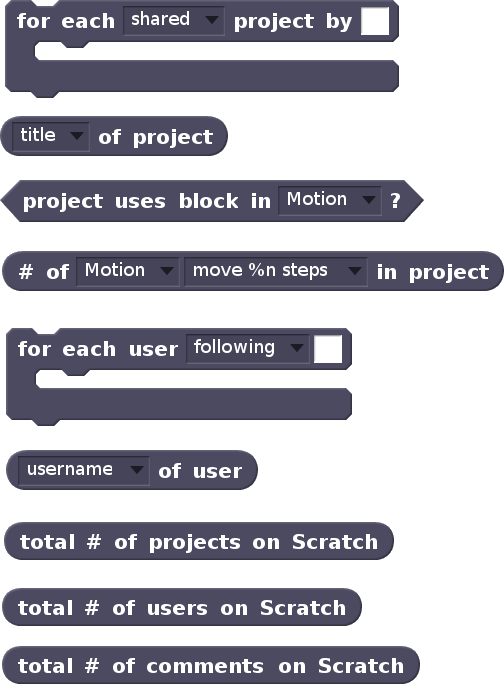}
\centering
\caption{Full palette of programming blocks made available through the \emph{Scratch Community Blocks} system.}
\label{fig:palette}
\end{figure}

\emph{Scratch Community Blocks} was designed as an extension to the Scratch language and was made available directly inside the Scratch editor. To use the systems, a user clicks on an ``Add Extension'' button in the ``More Blocks'' category. This shows a dialog box listing several available extensions that the user can pick from including the \emph{Community Blocks} system. When users choose the extension, a new palette of programming blocks, shown in Figure \ref{fig:palette}, appears in the programming editor.

Although the Scratch language does not meet all the criteria necessary to be an object-oriented language (e.g.,~there is no inheritance mechanism), the model of programming in Scratch can be described as object-centered. A sprite in Scratch can be thought of as an object with a number of properties that might include its position and orientation on the Scratch stage or a set of graphical ``costumes.'' Each sprite is also associated with a set of programmatic scripts that defines its behavior. \emph{Scratch Community Blocks} were modeled upon the \emph{accessor} or \emph{getter} method block for a Scratch sprite (shown in Figure \ref{fig:accessor}), which is a widely-used part of the core Scratch language. In an accessor block, the first dropdown menu lets the programmer select the property they are interested in; the second dropdown menu lets them choose a sprite from within the project. An accessor block can be embedded inside other blocks and will return data as determined by the parameters set through the two dropdown menus when executed.

\begin{figure*}[ht]
    \centering
    \begin{subfigure}[t]{0.32\textwidth}
        \centering
        \includegraphics[width=\columnwidth]{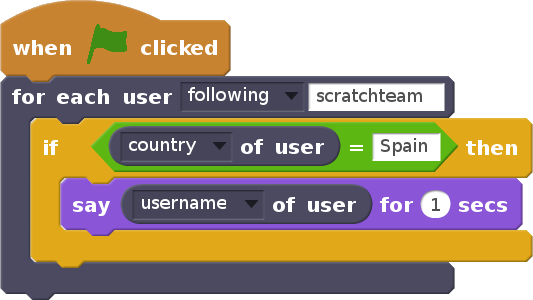}
        \caption{Script to filter followers by country.}
        \label{fig:sample-1}
    \end{subfigure}
    \hfill
    \begin{subfigure}[t]{0.32\textwidth}
        \centering
        \includegraphics[width=\columnwidth]{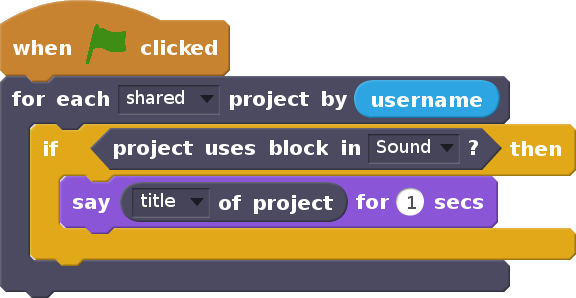}
        \caption{Script to filter projects by use of of blocks from the ``Sound'' category.}
        \label{fig:sample-2}
    \end{subfigure}
    \hfill
    \begin{subfigure}[t]{0.32\textwidth}
        \centering
        \includegraphics[width=\columnwidth]{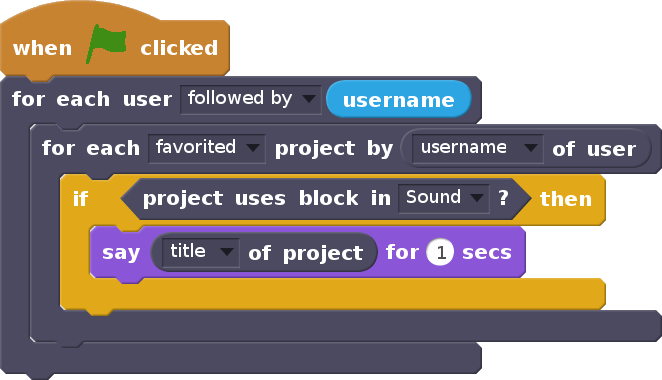}
        \caption{Script to find curated projects that use blocks from the Sound category.}
        \label{fig:sample-3}
    \end{subfigure}
    \caption{Simple code examples using \emph{Scratch Community Blocks} to show the system's design and functionality.}
    \label{fig:samples}
\end{figure*}

Instead of returning data on sprites within projects, \emph{Scratch Community Blocks} is designed to report data from across the Scratch online community. This model is also similar to the Object/Relationship Mapping (ORM) abstraction layer used to bridge an object-oriented language or development framework with an underlying relational database \cite{ambler_mapping_2000}. \emph{Scratch Community Blocks} provides data on the two most salient entities within the Scratch website and community: projects and users. Projects are the central site for interaction in Scratch and receive about 24\% of website traffic. User profiles are also seen as an important site for interaction and are the second most important locus with 7\% of website traffic.\footnote{Data for first quarter of 2016, from Google Analytics. Remaining traffic is to the front page, studios, and variety of other pages.} One design challenge was that while Scratch usernames are unique and widely known, projects are uniquely identified only by numeric IDs that are not surfaced in the Scratch user interface.

As a result, usernames are the only visible primary ``key'' in \emph{Scratch Community Blocks}. The user profile page in Scratch consists of data that the user shares about themself (e.g.,~``about me,'' ``what I'm working on''), as well as lists of projects shared, projects ``favorited'' or bookmarked, users followed (``followees''), and users who follow the user (``followers''). Using \emph{Scratch Community Blocks}, these lists are represented as a combination of ``for each'' loop blocks and context-sensitive accessor blocks that are intended to be used inside the loops. For example, to get the title of each project shared by a given user, the ``\texttt{title$\blacktriangledown$ of project}'' block needs to be placed inside the ``\texttt{for each shared$\blacktriangledown$ project by \underline{\hspace{0.5cm}}}'' as shown in the script in Figure \ref{fig:intro-script}. 

In \emph{Scratch Community Blocks}, the ``for each'' loop blocks represent queries to fetch:

\begin{itemize*}
    \item all shared projects by a user
    \item all favorited (bookmarked) project by a user
    \item all users who follow a given user
    \item all users who are followed by a given user
\end{itemize*}

For the project object type, three distinct context-sensitive accessor blocks are made available in the system. These blocks provide not only social metadata (e.g.,~number of comments on the project) but also code metadata (e.g.,~number of blocks of a certain type used in a given project, or whether a project uses blocks from a certain category). Specifically, the system includes a predicate block that returns a boolean \texttt{true} or \texttt{false} value depending on whether a project uses blocks from a certain category in Scratch (e.g.,~control, sound), a block that returns the number of instances of a specific block (e.g.,~``\texttt{wait \underline{\hspace{0.5cm}} seconds}''), and another block that can return the following social metadata on the project:

\begin{itemize*}
    \item title of the project
    \item description of the project
    \item number of ``love-its'' received by the project
    \item number of users who have bookmarked/favorited the project
    \item number of comments on the project
\end{itemize*}

For the user object type, there is only one accessor block, through which the following properties are made available:

\begin{itemize*}
    \item username of the user
    \item short public biography shared by the user (``about me'')
    \item country the user is from (publicly shared and self-reported)
\end{itemize*}

Although blocks reporting data on projects and users constitute most of the system, \emph{Scratch Community Blocks} also includes three reporter blocks, shown at the bottom of Figure \ref{fig:palette}, that return community-wide statistics: the total number of projects, users, and comments.

An open question for us was whether to add a set of primitives for drawing visualizations (e.g.,~bar charts or scatter plots). In the end, we did not include visualization primitives. Instead, we provided users with sample projects that used the ``pen'' primitives in Scratch to render several standard visualizations. As we show in the next section, children creating visualizations were frequently driven by their own artistic tastes and sensibilities and eschewed the more canonical approach that we showcased in the samples.

\begin{figure*}[ht]
    \centering
    \begin{subfigure}[t]{0.33\textwidth}
        \centering
        \includegraphics[width=\columnwidth]{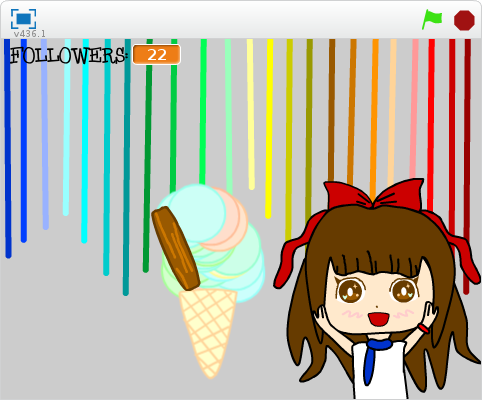}
        \caption{Visualization}
        \label{fig:example-3-viz}
    \end{subfigure}
    \hfill
    \begin{subfigure}[t]{0.65\textwidth}
        \centering
        \includegraphics[width=\columnwidth]{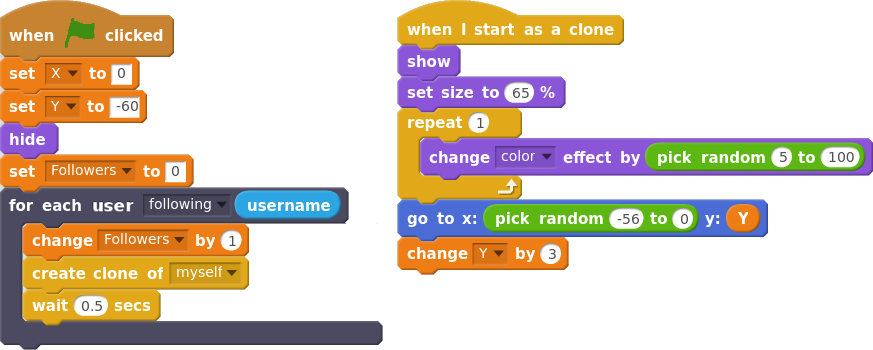}
        \caption{Code for the ``scoop'' sprite that creates the visualization through cloning the sprite.}
        \label{fig:example-3-code}
    \end{subfigure}
    \caption{Ice cream visualization project where the number of scoops on the cone is determined by the number of followers.}
    \label{fig:example-3}
\end{figure*}

\section{Illustrative Examples}

The best way to illustrate \emph{Scratch Community Blocks} is to show how it can be used. Toward that end, we present three simple examples that demonstrate the possibilities with the system. Figure \ref{fig:samples} shows the complete code of three small projects that demonstrate the operation of the system.
The first sample project, in Figure \ref{fig:sample-1}, demonstrates how one can filter, or search inside, the list of followers of a given user: the script iterates through the followers of the ``scratchteam'' user and ``says'' (in a speech bubble) the username of the followers who list Spain in the country field on their profile page.

Scratch has a number of blocks capable of producing sound, including blocks to play a recorded sound file, a musical note, or a drum. Within the Scratch system, these blocks are categorized as ``Sound'' blocks. The script shown in Figure \ref{fig:sample-2} will iterate over all the projects shared by a user and say the title of the projects that use blocks from the sound category. An important difference between this script and the one shown in Figure \ref{fig:sample-1} is the use of the ``\texttt{username}'' block that returns the username of the user currently viewing the project. This effectively provides customized  results for the Scratch user viewing the project.

In the final example in Figure \ref{fig:sample-3}, all the ``favorited'' projects of community members followed by the viewer are analyzed, and the title of the projects that contain blocks from the Sound category are presented. When run, this project will recommend projects that include sound or music to the viewer. These recommendations will have been socially curated by community members who are followed by the viewer.

\section{Field evaluation with children}

We believe that the most compelling illustrations of the system are projects created by real users in a realistic setting. In this section, we describe a series of projects, created by children, that serve as illustrations of how \emph{Scratch Community Blocks} can enable new pathways to learning data science and how it can make it possible for Scratch users to self-reflect on their learning and social participation. 

In February 2016, we invited users to beta test the \emph{Scratch Community Blocks}. Invitees were selected from a pool of active users on Scratch who had been in the community for at least 6 months and had shared at least 4 projects in the month preceding the selection. We looked only at project creation activity---we did not attempt to analyze the programming expertise or sophistication of these users. From this pool, 2,500 users were randomly selected and given access to the system in three phases. 
 
To help users understand how to program with the blocks, we provided users with a set of example projects, most similar to those shown in Figure \ref{fig:samples}, that were highlighted prominently in the landing page of the beta-test website. Additionally, the Scratch editor was modified to show a sidebar with documentation on the functionality of the blocks. As is always the case in Scratch, users had the ability to view and interact with other users' projects, including the corresponding source code, which is made available through the ``See Inside'' functionality in Scratch. Throughout the duration of the beta program, we also actively curated the website, placing projects that made interesting use of \emph{Scratch Community Blocks} on the landing page. Although the beta test took place on a separate website; the data accessed through \emph{Scratch Community Blocks} reflected what was there on the main Scratch website.

Self-reported gender data show that among the 2,500 users, 1,187 (47.5\%) were male, 1,184 (47.4\%) were female, 114 (4.6\%) identified themselves as other, and for 15 users, data was not available. The median age was 12, and the interquartile range was between 11 and 15. Over an approximately 4 month period, 721 of the invited users created 1,660 projects using the new blocks. These users created between 1 and 19 projects, with a mean of 2.3, median of 1, and standard deviation of 2.3. Toward the end of the beta test, we sent out a survey with 2 multiple choice and 8 open-ended questions to all the 2,500 invited users. 499 users responded. Additionally, we hosted a 3-hour workshop with 12 Boston-area Scratch users where we took detailed field notes and interviewed two participants. Finally, we interviewed 3 users from the online community over voice/video chat.

An analysis of the relative usage frequencies of the new blocks indicate that the most commonly used block was the one to get social metadata of projects (\texttt{title$\blacktriangledown$ of project}), followed by the one to iterate through projects (\texttt{for each shared$\blacktriangledown$ project by \underline{\hspace{0.5cm}}}). 86\% of the survey respondents rated the system as ``excellent'' or ``very good'', and 58\% reported that they spent at least an hour using it. A histogram showing the relative usage frequencies of the new blocks, and a full breakdown of the quantitative survey responses are included in the supplementary material for this paper.

Using a qualitative, open-ended coding process on the projects, interview transcripts, and comments on projects, we identified a variety of ways in which users were engaging in data science and reflection. We describe exemplars of these activities below. The projects presented are of higher quality than many of the other projects shared on the site, but they are not exceptional, and each is representative of a broader ``theme'' or group of projects and comments. Throughout the rest of this section, we use pseudonyms and altered usernames to refer to users.

\subsection{New pathways to data science}

Scratch users were able to use \emph{Scratch Community Blocks} to understand their world through computation and data. To that end, they created visualizations and representations of data as well as projects that tried to answer questions about social behaviors and learning activities of themselves and their peers.

\subsubsection{Representing and visualizing data}

\begin{figure}[h!]
\includegraphics[width=0.72\columnwidth]{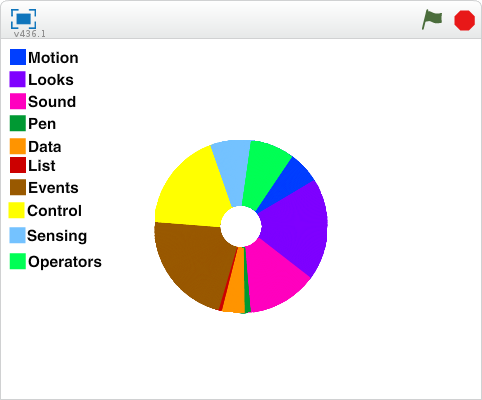}
\centering
\caption{Doughnut chart project showing distribution of all blocks ever used by a user over block categories.}
\label{fig:example-1}
\end{figure}

Jondroidous (13 years old) created a visualization project that prompted for a username and then visualized the distribution of block categories in all the projects shared by that user as a doughnut chart. An example of the visualization is shown in Figure \ref{fig:example-1}. When Jondroidous first created the project, there was a small bug in the code. As an illustration of the power of the social context provided by Scratch for learning and iteration, another user, Chewie184 (12 years old), remixed the project to fix the bug and improved the project by creating a progress bar.

As an illustration of how data analysis done by children may be more in line with the interests and desires of children than information visualizations created by adults, dashboards made using \emph{Scratch Community Blocks} were not restricted to ``standard'' visualization types such as pie charts, line graphs, etc. For example, AwesomeNemo (12 years old) made an ``ice cream visualization'' of a user's Scratch activity (shown in Figure \ref{fig:example-3}) where the number of scoops on the ice cream cone is determined by the number of followers. This project, as well as several other non-canonical visualization projects, made extensive use of Scratch's \texttt{clone} block that allows the creator of a project to programmatically create copies of sprites on the Scratch stage. In this particular project, a single sprite representing the scoop of ice cream was cloned repeatedly (number of repetitions = number of followers) to achieve the desired effect (Figure \ref{fig:example-3-code}).

To visualize one's creative and social activity on the site, AwesomeNemo made another project that consisted of a computationally generated island where various characteristics of the island were based on the viewer's activity in the community. She wrote in the description of the project on the website:

\begin{quote}
Your island will be generated on a basis of how many followers, shared projects, and favourited projects you have!: Beauty is how many stars your island has in the sky (favourited projects). Habitability is how many houses your island has (followers). Quality of life is how many trees your island has (shared projects).
\end{quote}

When asked about her motivation behind the island visualization during an interview with us, AwesomeNemo replied that she did not want to create a canonical visualization like a bar chart---rather, she wanted to explore more creative possibilities:

\begin{quote}
I was thinking about how I can make a project using more of the blocks, because [\ldots] I only used one in the ice cream one. So I was thinking about how I could use it to show statistics of how many followers and things you have, like not just in a bar chart.
\end{quote}

\subsubsection{Answering questions with data}

Many projects created with \emph{Scratch Community Blocks} answered questions on the activity of the user viewing these projects. In a sense, these were data science ``apps'' created by Scratch users for their peers.

\begin{figure}[h!] 
\includegraphics[width=0.85\columnwidth]{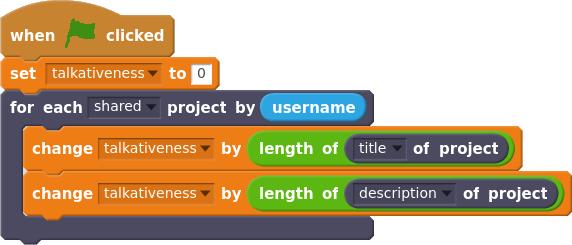}
\centering
\caption{Part of the code to calculate Scratch user's ``talkativeness'' (slightly simplified).}
\label{fig:example-talkative}
\end{figure}

A project by a 13-year-old titled ``How talkative are you?'' calculated the cumulative total length of the titles and descriptions of all the projects shared by the viewer of the project. It then added to this total the length of the username of the viewer and the length of the ``about me'' field on the the viewer's profile. The total was then presented to the viewer via a variable monitor widget on the Scratch stage (Figure \ref{fig:example-talkative} shows part of the code that did this analysis).
Other projects analyzed if any of the project viewer's followers mentioned them in their ``about me'' text, while others calculated the total number of followers the viewer's followers had. 

At the conclusion of the workshop, when we asked one of our participants (Sylvia) what other projects she would want to make with the blocks. She responded that before going on a vacation, she would like to make a project to find anyone in her social network on Scratch who lives in the place she's going, and then ask them for travel tips:

\begin{quote}
Let's say you want to go traveling, and you want to ask a Scratcher who's from there – so you could just search through all your followers, and you  can ask, ``Hey! How's Paris?''
\end{quote}

Embedded in this quote is evidence of Sylvia connecting her use of the new system with her existing knowledge (constructionism's continuity principle). Sylvia was, in essence, thinking about using \textit{Scratch Community Blocks} to express in a formal way what she already knew about the Scratch community. She knew that someone in her social network might be from France. She realized that, with \emph{Scratch Community Blocks}, she could access this information computationally to find those followers and connect with them.

\begin{figure}[h!]
\includegraphics[width=0.7\columnwidth]{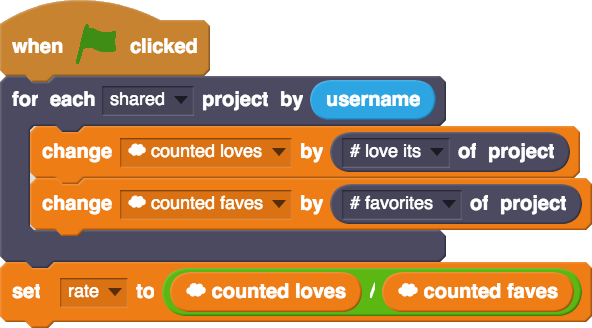}
\centering
\caption[Code to store and analyze data from multiple user trajectories using persistent Cloud Variables]{Code to store and analyze data from multiple user trajectories using persistent Cloud Variables (the $\vcenter{\hbox{\includegraphics{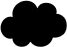}}}$ icon beside the variable names indicates that it is a Cloud Variable).}
\label{fig:example-4}
\end{figure}

There were also a few projects that attempted to answer questions about the Scratch community in general. Scratch supports persistent data structures through the \emph{Scratch Cloud Data} system created by Dasgupta \cite{dasgupta_surveys_2013}. Using this system, a variable in Scratch can be declared as persistent beyond the runtime of the project. Values set in the cloud variable are stored in a server, and the stored value is shared with all users accessing the project in question. Combining this system with \emph{Scratch Community Blocks}, TheCoder486 (12 years old) created a project that collected data about users who ran the project and stored metadata about the cumulative number of love-its and favorites for each set of projects analyzed. The numbers were stored in persistent variables, as shown in the code from the project in Figure \ref{fig:example-4}. With this simple approach, the project slowly built up a large sample dataset as more users accessed it. The question TheCoder486 was interested in was whether users tend to use the ``love it'' feature in Scratch more than the ``favorite'' feature. With the collected data as of date, his project suggests that love-its are 1.2 times more common than favorites.

\subsubsection{Incorporating data in game mechanisms}

Within the larger Scratch culture, games constitute a popular genre of projects created by community members. As the children using the \emph{Scratch Community Blocks} system were active members of the larger community and were already deeply embedded within the culture of Scratch, it was not unexpected to see a number of games (e.g.,~quizzes that asked questions about the viewer's past projects) created with \emph{Scratch Community Blocks} that incorporated data into their mechanisms. 

The 13-year-old who made the ``How talkative are you?'' project described earlier also made a ``data-driven'' doll dress-up game in which the viewer of the project has to pick and choose clothes and accessories for a virtual doll. Although there are a number of similar games on the Scratch website, what made this game different was the use of data about the player's participation in the Scratch community to determine the purchasing power of the player.

\begin{figure}[h!]
\includegraphics[width=0.72\columnwidth]{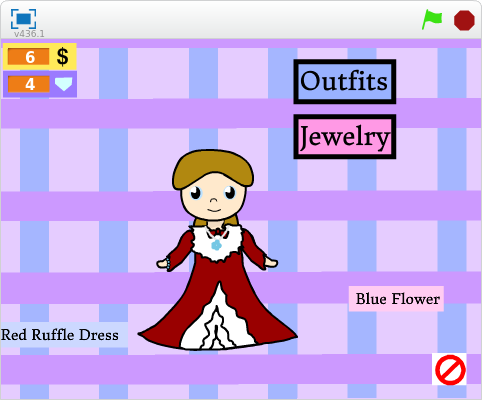}
\centering
\caption{Data-driven dress-up game in which social metrics are used to determine the project viewer's purchasing power.}
\label{fig:example-dressup}
\end{figure}

For each project shared, the player got one virtual dollar, and for each follower, a virtual diamond. The dollars and the diamonds could then be used to purchase clothes and jewelry for the virtual doll in the project (Figure \ref{fig:example-dressup}). When asked in an interview about how she came up with the idea for this project, the creator of the project responded by saying that she wanted to do something unique that incorporated her interest in making art on Scratch:

\begin{quote}
I was trying to think of something that somebody hadn’t done yet, and I didn’t see that. And also I really like to do art on Scratch and that was a good opportunity to use that and mix the two together.
\end{quote}

While games may not necessarily fall in the category of traditional data science tools, the children creating these projects were demonstrating fluency and creativity in ways in which they could put social and behavioral data to use. Additionally, as the example above shows, as children created projects, they connected to their own interests, identities, and aesthetic sensibilities.

\subsection{Self-reflection on learning and social participation}

Access to data and to data analytics tools created by their peers enabled Scratch users to self-reflect upon their own learning and social participation in Scratch. This self-reflection often happened in a public setting (in project comments), and in some cases, during interviews that we conducted. For example, on seeing the results of a pie-chart visualization of the relative proportion of block categories in their shared projects, a 15-year-old user commented, ``epic! looks like we need to use more pen blocks. :D''

In another example, during an interview with us, 14-year-old Alec said he realized that he usually focused on a certain type of block and did not really use others. We asked him if this realization meant that he would use the less frequently used blocks more in the future, and he replied affirmatively.

\begin{quote}
\textbf{Alec:} It made me think that I actually realized that sometimes I usually focus on a certain type of block. I usually make pen projects and I realized that I don't actually use sound or motion blocks that much.\\
\textbf{Interviewer:} Would you like to use them more, now that you know?\\
\textbf{Alec:} Yes, I probably would be.\\
\end{quote}

Not all self-reflection ended in positive feelings. After seeing the average number of ``love its'' on his projects, the 13-year-old author of the project that calculated this statistic left a sarcastic comment on his project: ``Average no. of loves---four. Well, that's not depressing at all :'(.''

However, by and large, most of the self-reflection we saw had a positive tenor. In a survey response, one user wrote:

\begin{quote}
I thought maybe I should expand my range of blocks that I would use commonly. I don't know what about it gave me this thought, but it did.
\end{quote}

In these projects, comments, and survey responses, we see children looking back at their own activities within the Scratch community, through data tools made by themselves and their peers to reflect on their participation in Scratch. By creating and sharing projects with \emph{Scratch Community Blocks}, Scratch users not only got a chance to reflect on their own activity, but also they enabled other members of the community---whoever viewed the project---to engage in the same reflective exercise.

\section{Implementation}

\begin{figure*}[ht!]
\includegraphics[width=1.0\textwidth]{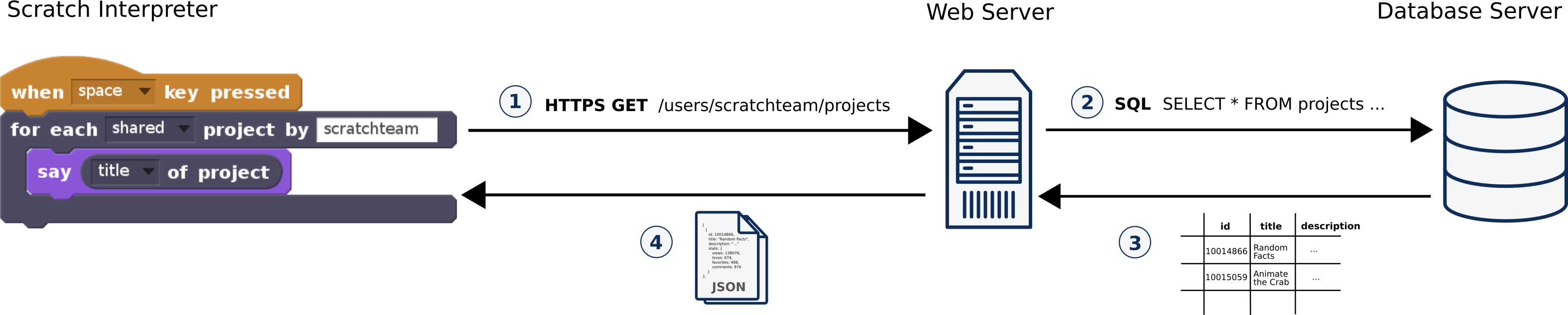}
\centering
\caption[System diagram for \emph{Scratch Community Blocks}]{System diagram for \emph{Scratch Community Blocks}. In step \circled{1}, the Scratch interpreter sends a HTTPS GET request to the Scratch API web server. In step \circled{2}, the API server translates the request URL into a SQL query and sends it to the database. The database sends back the data in step \circled{3}, which is then transformed into JSON and sent to the interpreter by the web server as a response to the original HTTPS request from step \circled{1}. Once the complete data is received, the interpreter starts the loop.}
\label{fig:system-diagram}
\end{figure*}

Though \emph{Scratch Community Blocks} is presented as an extension to the Scratch programmer, it is not written in JavaScript like standard Scratch extensions. Instead, it is implemented internally within the core Scratch code-base in ActionScript (the language Scratch is written in). This is due to the fact that the standard Scratch extension system does not allow for additions to the grammar of the language \cite{dasgupta_extending_2015}, and the design of \emph{Scratch Community Blocks} necessitates the new types of loops that are described in the previous section.

Internally, loop blocks in the \emph{Scratch Community Blocks} system send queries as HTTPS GET requests to the Scratch website API server. The API server parses the request, translates the request into an appropriate SQL statement, and then uses this SQL statement to query a database server. Results from the database are transformed into paginated JSON data and sent back to the interpreter as in response to the original HTTPS request. These responses are then parsed and stored for use by the query result accessor blocks. A diagram of this process is shown in Figure \ref{fig:system-diagram}. For code metadata result accessor blocks, an extra HTTPS request is sent to an online parser web service by the result accessor block itself. This online parser web service fetches the project being requested, parses the project code, and sends back the result as a response to the HTTPS request.

The Scratch interpreter, designed by Maloney et al.~\cite{maloney_scratch_2010}, implements a cooperative round-robin scheduler where every loop block in Scratch has a built-in yield. Thus, all runnable scripts get a chance to run every display cycle, providing the illusion of fine-grained concurrency. Blocks that request data from the server simply wait (i.e.,~pause execution of their script) while all other scripts (e.g.\ those which update the display) continue to run. When data arrives, execution proceeds in that script. This behavior is consistent with other blocks in Scratch that wait for something to complete before proceeding, such as the ``\texttt{wait \underline{\hspace{0.5cm}} seconds}'' or  ``\texttt{play sound \underline{\hspace{0.5cm}} until done}'' blocks.

A challenge with this approach is that Scratch was designed to be tinkerable and support ``live'' coding \cite{maloney_scratch_2010} with a minimal distinction between ``live'' and ``edit'' modes. This distinction is blurred to allow users to get a sense of what blocks or code can be used for. In \emph{Scratch Community Blocks}, the time taken by the website API to send back the full results over multiple pages can be non-negligible (i.e.,~on the order of a few minutes for the largest conceivable requests), and this lag can affect the perceived tinkerability of the system. As a partial workaround, we designed the system to cache results of queries and reuse the cached results when a given script runs a second time. This means that scripts will run quickly after an initial execution if its input parameters are unchanged. The trade-off is the possibility that a program written with \emph{Scratch Community Blocks} may operate with stale data. We felt that this trade-off was worth preserving the tinkerability that characterizes the Scratch language.

\section{Limitations}

Although \emph{Scratch Community Blocks} supported youth exploration of data, it was not without limitations. Some of its shortcomings were highlighted by our users through feedback in the forums and the survey. Common requests for enhancements included calls for more support materials (documentation and sample projects), as well as for a faster and more responsive system. Additionally, we identified several difficulties faced by users that point toward shortcomings and limitations of the current design.

\begin{figure}[h!]
\includegraphics[width=0.82\columnwidth]{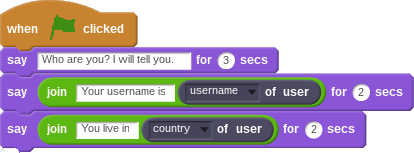}
\centering
\caption{Code from user-created project that illustrates how the context-sensitive accessor block is incorrectly used.}
\label{fig:misconception-1}
\end{figure}

One common misconception (observed in project shared by 19\% of users) involved the use of the context-sensitive accessor blocks. Context-sensitive blocks can be thought of as being similar to variables that have limited scope. In the case of the \emph{Scratch Community Blocks}, most accessor blocks are not ``valid'' outside of their associated loops. This pattern of usage was more common with the accessor block for user data. In most of these cases, as in the example shown in Figure \ref{fig:misconception-1}, it appears that the creator of the project expected the block to return data about the user currently viewing the project.

Context-sensitive blocks are a general problem with blocks-based languages. For example, in Scratch, text input from a user is stored in an ``\texttt{answer}'' block that is valid only if the programmer had previously prompted the user for input. The current design of \emph{Scratch Community Blocks} uses these context-sensitive blocks prominently, making this problem particularly salient. Currently, other than showing an error message---an approach Scratch has explicitly avoided supporting \cite{maloney_scratch_2010}---there is no obvious or easy workaround to this problem.

Another source of difficulty among users was in correctly scaling visualizations. For example, a very active user created a ``cupcake visualization'' where the size of the cupcake depended on the number of projects shared by the viewer. If the project was viewed by another user with a small number of shared projects, the cupcake image would be scaled to a size that made it barely visible. The bug alluded to earlier in the doughnut chart visualization example in Figure \ref{fig:example-1} was caused by the fact that the sections of the doughnut were scaled by the number of blocks used by the project's creator. As a result, if a user who had used fewer blocks viewed the project, the visual did not form a complete circle. If a user who had used more blocks viewed it, the sections on the chart would overflow and draw over previously drawn elements. This issue can be seen as a consequence of our decision not to include visualization primitives. If we had included visualization primitives with inbuilt mechanisms for scaling, our users would not have encountered this particular difficulty. 

Additionally, it was clear in several cases that a project creator expected data accessed using the blocks to be updated in real time. This was commonly seen with the overall statistics blocks, as they were embedded within loops in some cases, indicating that the author of the project expected new values in different iterations of the loop. Unfortunately, real-time updating is not supported by the underlying architecture of the system.

Finally, while we found many examples of successful uses of the system, we have little data about invited users who were unsuccessful in using the system. In this sense, our evaluation may be biased toward users who were successful in using the system on their own, at least in the bulk of the data drawn from outside of the workshop. This is a common challenge in studies of informal learning online. Though a significant portion of the users that we invited were never active on the website, we do not know if this inactivity was due to difficulties in using the system or if they were busy or uninterested. Future studies in more controlled contexts (e.g.,~ in workshops with researchers present) may shed more light on the overall usability of the system.

\section{Conclusion}

In this paper, we presented the motivation and design of \emph{Scratch Community Blocks}. We explained how the system is designed to allow children to engage directly in processes of data analysis normally practiced only by adults in order to satisfy our twin goals of promoting new pathways to learning about data analysis and promoting reflection. We demonstrated, through a series of examples and case studies, how \emph{Scratch Community Blocks} both engages children in visualizing, representing, and answering questions with data in new ways and also supports self-reflection on learning and social participation. Finally, we presented a discussion of the limitations of the current design based on problems that users encountered.

As a system designed to support a constructionist approach to data science, \emph{Scratch Community Blocks} enables learners to learn by designing and building projects that help them answer questions they are interested in and to represent and make use of data in ways that speak to their styles and identities. Learners can also share their work with their peers. These shared creations are not only viewable by others but can also be used by peers to reflect on their experiences and learning. 

Among the related systems that informed our design, \emph{Scratch Community Blocks} is a unique design in several senses. The system not only fosters data-mediated reflection on learning, it also enables users to construct their own visualizations and analyses in ways that are often very different from canonical forms of data representation and analysis. We believe that this unique combination of affordances is potentially useful in a wide range of personal informatics contexts.

For educators, our system can be seen providing a unique pathway into data science. While many introductory toolkits for data science education focus on data sets that may or may not be of particular relevance to learners, users of \emph{Scratch Community Blocks} analyze data that is about themselves and that has been created by themselves.
Although Scratch is an informal context for learning, we believe that a system such as ours has merits and could be used in formal settings as well. As described by Brennan \cite{brennan_best_2013}, an interest-driven and exploration-focused approach like the one supported by our system might be utilized to enable learners to see ``entry points and trajectories of participation'' in an existing disciplinary realm, such as data science.

Although the projects created by youth that we have presented can appear superficially different from the kinds of analysis completed by educational researchers and learning scientists, children are using \emph{Scratch Community Blocks} to write computer programs to ask and answer question using data about their own activities and learning. In his article ``Teaching Children to be Mathematicians vs.~Teaching About Mathematics'' \cite{papert_teaching_1971}, Papert showed how Logo could offer children a space to use and engage with mathematical ideas in creative and personally motivated ways. This, he argued, enabled children to go beyond knowing about mathematics to ``doing'' mathematics, as a mathematician would. In a similar fundamental sense, \emph{Scratch Community Blocks} allows children to \emph{do} data science, and not just know about it.

In his book \emph{Mindstorms: Children, Computers, and Powerful Ideas}, Papert presented the idea of the ``child as the epistemologist,'' where children, through construction and reflection, not only learn new and powerful ideas but also think about their own thinking and learn about their own learning \cite{papert_mindstorms:_1980}. Inspired by Papert's vision, we feel that the greatest promise of \emph{Scratch Community Blocks} and its approach is that, in enabling children to understand and analyze their own learning and social participation, it may not only help children become data scientists and analysts but also encourage them to become epistemologists as well.

\section{Acknowledgments}

This paper is based on the first author's PhD dissertation, and some of the text and images in the paper appear in the thesis. We would like to thank members of the Scratch community for their time and for inspiring us with their creative work. We would also like to acknowledge Mitchel Resnick, Natalie Rusk, Hal Abelson, John Maloney, and our anonymous reviewers for their support and thoughtful feedback. Financial support for this work came from the National Science Foundation (grants DRL-1417663 and DRL-1417952).

\balance
\bibliographystyle{SIGCHI-Reference-Format}
\bibliography{references}
\end{document}